\documentclass[10pt,english,aps,prx,twocolumn,superscriptaddress,floatfix]{revtex4-2}

\usepackage[utf8]{inputenc}

\usepackage[T1]{fontenc}
\usepackage{babel}
\usepackage{amsmath,amssymb,mathtools}
\usepackage{bm}
\usepackage{graphicx}
\usepackage{braket}
\usepackage{xspace}
\usepackage{xcolor}
\usepackage{physics}
\usepackage{bbm}
\usepackage{comment}
\usepackage{silence}
\usepackage{hyperref}

\newcommand{\om}{\omega}

\newcommand{\Fop}{\mathcal F}

\begin{document}
\title{Chiral Parafermions in Floquet \texorpdfstring{$\mathbb{Z}_N$}{ZN} Topological Order}
\author{M. Will}
\thanks{These authors contributed equally.}
\affiliation{\mbox{Physics Department, TUM School of Natural Sciences, Technical University of Munich, 85748 Garching, Germany}}
\affiliation{Munich Center for Quantum Science and Technology (MCQST), Schellingstr. 4, 80799 M{\"u}nchen, Germany}
\affiliation{Department of Physics, Harvard University, Cambridge, Massachusetts 02138, USA}
\author{H. Schmid}
\thanks{These authors contributed equally.}
\affiliation{\mbox{Physics Department, TUM School of Natural Sciences, Technical University of Munich, 85748 Garching, Germany}}
\affiliation{Munich Center for Quantum Science and Technology (MCQST), Schellingstr. 4, 80799 M{\"u}nchen, Germany}
\author{J. Egger}
\affiliation{\mbox{Physics Department, TUM School of Natural Sciences, Technical University of Munich, 85748 Garching, Germany}}
\affiliation{Munich Center for Quantum Science and Technology (MCQST), Schellingstr. 4, 80799 M{\"u}nchen, Germany}
\author{A. Gammon-Smith}
\affiliation{\mbox{School of Physics and Astronomy, University of Nottingham, Nottingham, NG7 2RD, UK}}
\affiliation{Centre for the Mathematics and Theoretical Physics of Quantum Non-Equilibrium Systems, University of Nottingham, Nottingham, NG7 2RD, UK}
\author{M. Knap}
\affiliation{\mbox{Physics Department, TUM School of Natural Sciences, Technical University of Munich, 85748 Garching, Germany}}
\affiliation{Munich Center for Quantum Science and Technology (MCQST), Schellingstr. 4, 80799 M{\"u}nchen, Germany}
\author{F. Pollmann}
\affiliation{\mbox{Physics Department, TUM School of Natural Sciences, Technical University of Munich, 85748 Garching, Germany}}
\affiliation{Munich Center for Quantum Science and Technology (MCQST), Schellingstr. 4, 80799 M{\"u}nchen, Germany}

\begin{abstract}
Floquet topological order (FTO) extends topology beyond equilibrium, giving rise to phenomena, such as the dynamical transmutation of anyons. Here, we realize $\mathbb{Z}_N$ FTO in a generalized Kitaev honeycomb model with clock degrees of freedom using a three-step periodic drive. The driven system admits an emergent description in terms of constrained parafermions coupled to conserved $\mathbb{Z}_N$ gauge fields. At the boundary, the drive generates chiral parafermionic Floquet edge modes. We resolve their propagation through correlation functions, characterize their fractional quantum-information transport through the radical chiral unitary index, and use their dynamics to probe exchange statistics. In the bulk, we explore the dynamical anyon transmutation with a non-equilibrium loop order parameter. 
At a Floquet sweet spot, where the drive becomes Clifford, these phenomena can be accessed using stabilizer simulations. Short-time matrix-product-state simulations away from this point show that the chiral edge response persists at accessible times. Our results establish $\mathbb{Z}_N$ Kitaev models as a setting for exploring intrinsically non-equilibrium topological phenomena on qudit-based quantum processors.
\end{abstract}

\date{\today}
\maketitle
\section{Introduction}
Identifying controllable many-body systems that realize exotic non-equilibrium phases remains a central challenge in condensed matter physics. Among these, topologically ordered phases stand out due to their fractionalized anyonic excitations \cite{wen_topological_1990,kitaev_fault-tolerant_2003,nayak_non-abelian_2008} and long-range entanglement \cite{kitaev_topological_2006,levin_detecting_2006}. Recent experiments have demonstrated the preparation and control of topologically ordered states on programmable quantum processors, offering a window into their complex many-body dynamics \cite{satzinger_realizing_2021,semeghini_probing_2021,bluvstein_quantum_2022,andersen_non-abelian_2023,iqbal_topological_2024,berthusen_experiments_2024,xu_non-abelian_2024,iqbal_qutrit_2025,GammonSmith2026} and providing a promising route towards quantum error correction \cite{bluvstein_logical_2024,acharya_quantum_2025}.

Periodic driving can realize non-equilibrium topology \cite{Kitagawa2010,lindner2011floquet,Jiang2011}, such as anomalous Floquet insulators with dynamical chiral edge modes \cite{rudner_anomalous_2013,nathan_anomalous_2019}. This anomalous Floquet physics extends naturally to topologically ordered phases \cite{Po2016_chiral,Potter2017,po_radical_2017,zhang_bulkboundary_2023,petiziol_nonabelian_2024}.
A natural testbed for $\mathbb{Z}_2$ Floquet topological order (FTO) is the periodically driven spin-$\frac{1}{2}$ Kitaev honeycomb model \cite{kitaev_anyons_2006,po_radical_2017,fulga_topology_2019,molignini_crossdimensional_2021,kishony_topological_2025}, recently realized on a quantum processor \cite{will_probing_2025}. Its emergent Majorana excitations acquire nontrivial dynamics under driving. In the bulk, the drive permutes electric and magnetic anyons, realizing a dynamical transmutation between different anyon superselection sectors. At the boundary, the chiral propagation of Majorana edge modes generates a quantized flow of quantum information with fractionalized information content \cite{Po2016_chiral,Duschatko2018}. FTO is expected to persist as a prethermal phase \cite{po_radical_2017,Abanin2015,Bukov2015,PhysRevLett.116.120401,Abanin2017,Else2017} and may be stabilized by many-body localization (MBL) \cite{Po2016_chiral}.

\begin{figure*}[t]
    \centering
    \includegraphics[width=\linewidth]{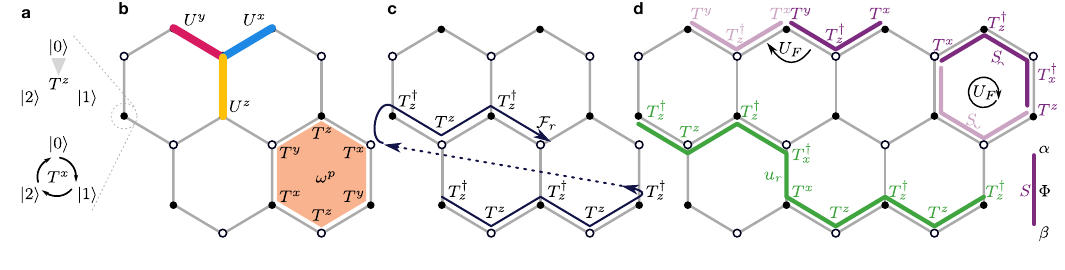}
    \caption{\textbf{$\mathbb{Z}_N$ Floquet Kitaev honeycomb model.} \textbf{a} Actions of the clock operators on the local qudit state (illustrated for the $\mathbb{Z}_3$ qutrit case). $T^z$ is diagonal in the local qudit basis, while $T^x$ shifts between the basis states.
    \textbf{b} The Floquet operator $U_F = U^zU^yU^x$ conserves the plaquette fluxes $W_p=\omega^p,\, p \in \mathbb{Z}_N $, see Eq.~\eqref{eq:plaquette_operator}.
    \textbf{c} 
    Schematic Fradkin-Kadanoff-type contour which maps clock operators to constrained operators with parafermionic exchange statistics on distinct sites.
    \textbf{d} Conserved $\mathbb{Z}_N$ gauge link $u_r$ (green) is associated with $z$-bonds. Strings of clock operators can be mapped to parafermions ($\alpha_r,\beta_{r'}$) at the endpoints, connected by a conserved string ($\Phi$). At the edge, strings evolve along the contour, whereas in the bulk they circle around a plaquette.}
    \label{fig:fig1}
\end{figure*}

A key question is how FTO extends beyond the $\mathbb{Z}_2$ setting. In equilibrium, Barkeshli et al.~\cite{barkeshli_generalized_2015} introduced a $\mathbb{Z}_N$ generalization of the Kitaev model, which has since been studied extensively \cite{Eck_Kitaev_N,Chen2024,Yang2025}, where spin-$\frac{1}{2}$ operators are replaced by clock spin operators of local dimension $N>2$, see Fig.~\ref{fig:fig1}a. The resulting model hosts equilibrium $\mathbb{Z}_N$ topological order \cite{Levin2005,Schulz_2012,Dong2017,Kaustubh2024} characterized by abelian anyons with $\frac{2\pi}{N}$ exchange statistics as well as parafermionic edge modes, which have also attracted interest in one-dimensional settings and coupled two-dimensional arrays \cite{lindner_fractionalizing_2012,clarke_exotic_2013,fendley_parafermionic_2012,Mong2014,alicea_topological_2016,Stoudenmire2015,Vaezi2015,Son2018,Benhemou2023,Wei2015,Calzona2018,Vinkler2017}. 
Crucially, the generalized Kitaev model for $N>2$ is no longer exactly solvable due to its interacting parafermionic nature. Furthermore, a recent experiment has realized the $\mathbb{Z}_3$ toric code on a trapped-ion quantum processor \cite{iqbal_qutrit_2025}, demonstrating that such $\mathbb{Z}_N$ phases are becoming experimentally accessible on qudit platforms and qudit encodings based on qubits.

Here, we study a three-step periodic drive of a generalized Kitaev model built from clock matrices to realize $\mathbb{Z}_N$ FTO. Using a Fradkin-Kadanoff-type mapping, we rewrite the model in terms of constrained parafermions coupled to conserved $\mathbb{Z}_N$ gauge fields. These parafermions obey generalized exchange relations, extending the 
anticommutation relations of Majorana fermions to $\mathbb{Z}_N$ degrees of freedom. At a Floquet sweet spot, the dynamics reduce to a Clifford circuit, which describes the hopping of these constrained parafermions and allows for stabilizer simulations. In the bulk, the Floquet evolution exchanges generalized electric and magnetic anyons $e\leftrightarrow m$, giving rise to an anyonic time crystal \cite{will_probing_2025}, an anyonic analogue of discrete time crystals \cite{Khemani2016,else2016floquettimecrystals,von_Keyserlingk_2016}.
At the edge, the drive generates chiral parafermion transport. We develop experimentally accessible protocols to directly probe these signatures of FTO. Specifically, we track the parafermionic edge mode using correlation functions. Away from the Clifford point, we perform short-time matrix-product-state (MPS) simulations, which are limited by rapid entanglement growth. 
We further characterize the quantized edge information transport through the chiral mutual information, whose growth yields the radical chiral unitary index \cite{Duschatko2018}. Lastly, we show how the Floquet dynamics can be used in an interferometric protocol to probe parafermionic exchange statistics. Together, these results establish the generalized Kitaev model as a concrete route towards realizing and probing $\mathbb{Z}_N$ FTO on programmable qudit platforms. 

\section{Model}
We construct a $\mathbb{Z}_N$ generalization of the Floquet Kitaev model 
with $N$-level local degrees of freedom. These degrees of freedom are located on the vertices of a honeycomb lattice, with the three bond orientations labeled by $x,y,z$, see Fig.~\ref{fig:fig1}b. During one Floquet cycle, the three bond types are applied sequentially, giving the stroboscopic evolution $U_F = U^zU^yU^x$
with 
\begin{align}
    U^{\alpha}
=
\prod_{\langle k,l\rangle_\alpha}
\exp\!\left[
iJ
\sum_{n=1}^{N-1}
c_n
\left(T_k^\alpha T_l^\alpha\right)^n
\right].
    \label{eq:Floquet pulses exp}
\end{align}
Here, $\langle k,l\rangle_\alpha$ denotes nearest-neighbor pairs along the bond $K_{kl}=T_k^\alpha T_l^\alpha$ with orientation $\alpha\in\{x,y,z\}$, and $J$ is the coupling strength. We use open boundary conditions throughout. The clock matrices $T^\alpha$ are the $N\times N$ generalizations of the Pauli operators with $(T_i^\alpha)^N=1$.
In the local qudit basis $\ket{n}$, 
they act as
\begin{align}
T^z\ket{n}=\omega^n \ket{n}
   \quad 
T^x\ket{n}=\ket{n+1 \, \mathrm{mod}\,N},
\end{align}
with $\omega=e^{2\pi i/N}$, as shown for the $\mathbb{Z}_3$ case in Fig.~\ref{fig:fig1}a.
The third operator is fixed by $T^y=\zeta(T^z T^x)^\dagger$, where $\zeta = 1$ for odd $N$ and $\zeta = e^{i\pi/N}$ for even $N$. They satisfy the on-site commutation relations $T^\alpha_iT^\beta_i=\omega^{\epsilon_{\alpha \beta}} T^\beta_i T^\alpha_i$ with $\epsilon_{xy}=\epsilon_{yz}=\epsilon_{zx}=1$, while operators on different sites commute. 

Importantly, the coefficients $c_n=-\frac{\pi}{2N}\sin^{-2}\left(\frac{\pi n}{N}\right)$ in the drive are chosen such that at $J=1$ all $U^\alpha$ are Clifford gates, see App.~\ref{app:clifford}. This is the $\mathbb{Z}_N$ extension of the $\mathbb{Z}_2$ Clifford point to arbitrary $N$, and allows us to understand the evolution analytically. 

Away from the Clifford point, we expect that weak perturbations may give rise to a long-lived prethermal phase \cite{Abanin2015,Bukov2015,Abanin2017,PhysRevLett.116.120401,Else2017}. Strong disorder could further suppress heating and potentially stabilize FTO through MBL \cite{Po2016_chiral}, although asymptotically stable MBL in two dimensions remains unsettled \cite{DeRoeck2017}.

Analogously to the $\mathbb{Z}_2$ case the Floquet evolution conserves the $\mathbb{Z}_N$ plaquette operator
\begin{align}
    W_p = \prod_{\langle i,j\rangle \in p } K_{ij} = \left(\omega \zeta^{-6}T^y_1T^z_2T^x_3T^y_4T^z_5T^x_6\right)^\dagger,
    \label{eq:plaquette_operator}
\end{align}
where $K_{ij}$ are ordered clockwise around a plaquette starting with the upper left $x$-bond. 
Its eigenvalues $W_p=\omega^p$ with $p\in\mathbb{Z}_N$ measure the quantized flux through the plaquette. 

\section{Parafermions}

\subsection{Fradkin-Kadanoff mapping}

We next map the clock model to parafermions, which generalize Majorana fermions to $\mathbb{Z}_N$ degrees of freedom. In general, parafermions obey the exchange algebra
\begin{align}
(\alpha_k)^N=1, \qquad
\alpha_k\alpha_{k'}=\omega^{\mathrm{sgn}(k'-k)}\alpha_{k'}\alpha_k,
\end{align}
where $\{k\}$ is a totally ordered set.
We introduce a Fradkin-Kadanoff-type transformation that maps the clock operators onto \textit{constrained} parafermions \cite{fradkin1980disorder1}. Following the $\mathbb{Z}_2$ mapping of Refs.~\cite{ChenZohar2008,PhysRevLett.98.087204}, the exchange statistics are encoded in nonlocal strings, rather than arising from local parton operators as in Kitaev's Majorana construction \cite{kitaev_anyons_2006}.
The resulting operators obey parafermionic exchange statistics on distinct
physical sites, while retaining a noncanonical on-site algebra. In spirit, our construction is analogous to hard-core bosons which behave as bosons beyond the on-site scale \cite{10.1143/PTP.16.569}.

Since the honeycomb lattice is bipartite, we label the two sublattices in white and black and choose a row-by-row ordering of the lattice sites, as illustrated in Fig.~\ref{fig:fig1}c. 
We associate with each site $r$ the string operator
\begin{align}
    \Fop_r&=\prod_{r' \prec r}\left(T^z_{r'}\right)^{s_{r'}},    
    \quad s_{r'} =
    \begin{cases}
        +1, & r' \in \text{white},\\
        -1, & r' \in \text{black}.
    \end{cases}
    \label{eq:string}
\end{align}
Here, $r'\prec r$ denotes that $r'$ precedes $r$ along the chosen row-by-row ordering. Thus, $\Fop_r$ contains $T^z$ on every preceding white site and ${T^{z}}^\dagger$ on every preceding black site. Next, we define four parafermions per physical site
\begin{align}
A_{r}&=\Fop_{r}T^x_{r},
\qquad
&\widetilde A_{r}&=\Fop_{r}{T^x_{r}}^\dagger, \\
B_{r}&=\Fop_{r}{T^z_{r}}^\dagger {T^x_{r}}^\dagger, 
\qquad
&\widetilde B_{r}&=\Fop_{r}T^x_{r} T^z_{r}.
\label{eq:app_parafermions}
\end{align}
The four operators are not independent and preserve the number of physical degrees of freedom, as reflected in the local identity
$A_{r}\widetilde A_{r} = B_{r} \widetilde B_{r}$.
Unlike for fermions, exchanging two parafermion operators produces an ordering dependent phase rather than a fixed minus sign. Their exchange algebra is in this case specified by a snake ordering, which is distinct from the spatial row-by-row ordering.
This parafermionic snake winds once forward through the lattice and once backward, ordering the four parafermions associated with each physical site as
\begin{align*}
&\widetilde B_1 \rightarrow A_1\rightarrow B_2 \rightarrow \widetilde A_2\rightarrow \ldots \rightarrow B_L\rightarrow \widetilde A_L \\
&\rightarrow\widetilde B_L\rightarrow A_L\rightarrow \ldots \rightarrow\widetilde B_2\rightarrow A_2 \rightarrow B_1\rightarrow \widetilde A_1.
\end{align*}
Here $1, \dots, L$ label physical sites according to the row-by-row ordering, with the first site chosen on the black sublattice and the last on the white sublattice. We use $\alpha\prec\beta$ to denote that parafermion $\alpha$ appears before $\beta$ along the snake. For any two parafermions $\alpha$ and $\beta$ on distinct physical sites, the ordering along the parafermionic snake then determines their exchange relation 
\begin{align}
    \alpha\beta = \om \beta \alpha \qquad  \text{if } \alpha\prec\beta,
\end{align}
where $\alpha,\beta\in\{A_r,B_r,\widetilde A_r,\widetilde B_r\}$.
Their physical on-site algebra differs from that of canonical parafermions, reflecting the local constraint. Nevertheless, on distinct sites the operators obey parafermionic exchange statistics, which will allow us to interpret the Floquet steps in terms of parafermionic bilinears.

Under the transformation, the three types of bonds become
\begin{align}
    T^x_{r} T^x_{r+1} &= \widetilde B_{r} \widetilde A_{r+1}^\dagger  
    \label{eq:xx-bond PF}\\
    T^y_{r}T^y_{r+1} &= \om^{-1} \zeta^{2} A_{r}^\dagger B_{r+1}
    \label{eq:yy-bond PF}\\
    T^z_{r+L_x}T^z_{r} &= B_{r+L_x}^\dagger \widetilde A_{r} u_{r}
     \label{eq:zz-bond PF}, 
\end{align}
with $L_x$ being the number of sites in the horizontal direction. The $x,y$-bonds map directly onto parafermionic bilinears, while the vertical $z$-bonds acquire an additional nonlocal link operator 
\begin{align}
    u_{r} = \omega B_{r}^\dagger \widetilde A_{r+L_x},
\end{align}
shown in Fig.~\ref{fig:fig1}d. Importantly, $u_{r}$ commutes with $U_F$ and is therefore a conserved $\mathbb{Z}_N$ gauge-link variable. The construction generalizes the $\mathbb{Z}_2$ case, where Majoranas are coupled to conserved $\mathbb{Z}_2$ gauge fields \cite{ChenZohar2008}.
The flux is encoded in products of neighboring gauge links
\begin{equation}
W_p=\zeta^4u_{r}^\dagger u_{r+2}.
\label{eq:plaquette_u}
\end{equation}
Detailed derivations can be found in App.~\ref{SI:parafermion_mapping}.

\subsection{Parafermion dynamics}
We now turn to the dynamics of the emergent parafermions. An open string of bond operators along a path $\gamma_{r,r'}$ can be written as
\begin{align}
S_{rr'}=\prod_{\langle k,l\rangle\in\gamma_{r,r'}}K_{kl}
=\alpha_{r}\beta_{r'}\Phi_{\gamma_{r,r'}},
\label{eq:PF string}
\end{align}
corresponding to two parafermions $\alpha_{r}$ and $\beta_{r'}$ at its endpoints. All phases are absorbed into the connecting gauge string $\Phi_{\gamma_{r,r'}}$, see Fig.~\ref{fig:fig1}d. Further details and examples are given in App.~\ref{SI:parafermion_mapping}. For a state with definite parafermion-pair occupation $n_{\rm PF}$, the string has expectation value $\langle S_{rr'}\rangle=\omega^{n_{\rm PF}}$. The dynamics of the emergent parafermions can therefore directly be inferred from the evolution of the corresponding clock strings.

At the Clifford point $J=1$, this motion can be tracked analytically, since string operators remain string operators under the evolution. Under each pulse, the string elongates on one side while being shortened on the other. This can be understood in the parafermion language, where the drive realizes the braid operator in this limit~\cite{lindner_fractionalizing_2012, clarke_exotic_2013,hutter_quantum_2016}. Explicitly, each bond pulse becomes (up to a global phase)
\begin{align}
 U^{\alpha}_{kl}= \frac{1}{\sqrt{N}}\sum_{n=0}^{N-1}\omega^{n(n+N)/2}\left(K_{kl}\right)^{n},
 \label{eq:braid operator}
\end{align}
where $K_{kl}\propto\alpha_k\beta_l^\dagger$ are the parafermion bilinears of Eqs.~\eqref{eq:xx-bond PF}-\eqref{eq:zz-bond PF}.

When time evolving $S_{rr'}$, the gauge string $\Phi_{\gamma_{rr'}}$ commutes with the drive, such that only the endpoint parafermions evolve. Each pulse acts on the parafermions as a nonlinear hopping operator. Namely, on one end the parafermion hops across the bond, while at the other end of the string it acquires additional parafermionic factors. The latter can be recombined into the invariant gauge string, effectively extending it at one end, while shortening it at the other. The time-evolved string operator therefore keeps the form of parafermions at its endpoints connected by a gauge string, 
see Eq.~\eqref{eq:PF string} and App.~\ref{sec:pf dynamics}
for details.

Parafermions acquire different dynamics in the bulk and at the boundary, see Fig.~\ref{fig:fig1}d. In the bulk, the strings periodically encircle plaquettes, and the corresponding parafermions perform localized periodic orbits. In contrast, the strings at the edge are translated by one unit cell during each Floquet period, 
corresponding to chirally propagating parafermions along the edge. This realizes a FTO phase \cite{Po2016_chiral,po_radical_2017,Potter2017}.

\begin{figure*}[t]
    \centering
\includegraphics[width=\linewidth]{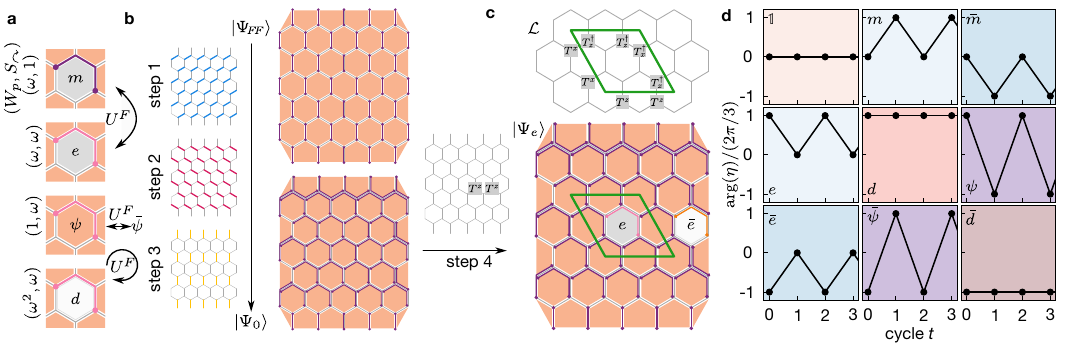}
    \caption{\textbf{$\mathbb{Z}_3$ Floquet anyon transmutation.} \textbf{a} Anyons are labeled by the eigenvalues of the plaquette operator $W_p$ and the half-plaquette string $S_{\curvearrowright}$, corresponding to the magnetic flux $m$ and the parafermion $\psi$, respectively. The Floquet drive exchanges the electric and magnetic anyons, $\psi$ and its antiparticle, and leaves the dyon $d$ invariant. Representative $\mathbb{Z}_3$ anyons are shown; antiparticles are omitted. \textbf{b} Protocol for preparing an anyon excitation and probing its Floquet dynamics. The flux-free state ($\ket{\Psi_{\mathrm{FF}}}$) is transformed into plaquette-diagonal parafermion pairings ($\ket{\Psi_0}$). A string of $T^z$ operators creates an $e$-$\overline{e}$ pair. \textbf{c} Loop operator $\mathcal{L}$ encircling the $e$ anyon. The green path indicates the trajectory of the braiding anyon. 
    \textbf{d} Phase of the non-equilibrium loop order parameter $\eta$, obtained from stabilizer simulations for the $\mathbb{Z}_3$ model at $J=1$. Panels with the same background color correspond to anyon pairs which are transmuted into one another.}
    \label{fig:fig2}
\end{figure*}

\section{Floquet anyon transmutation}
To investigate the bulk dynamics further, we consider a string $S_{\curvearrowright}=K_{12}K_{23}K_{34}$ with associated occupation $n_{\rm PF}$ which spans one half of a plaquette carrying flux $p$, as shown in Fig.~\ref{fig:fig1}d. Under one Floquet period, the string is mapped onto the complementary half of the plaquette
\begin{align}
    {U_F}^\dagger S_{\curvearrowright} U_F =S_{\rotatebox[origin=c]{180}{$\curvearrowright$}}= S^\dagger_{\curvearrowright} W_p,
\label{eq:S_timeevolved}
\end{align}
where $S_{\rotatebox[origin=c]{180}{$\curvearrowright$}}=K_{45}K_{56}K_{61}$. After a second Floquet period, the original string is recovered, giving rise to period-two dynamics. The string occupation therefore transforms over one period as
$n_{\rm PF}
\rightarrow p - n_{\rm PF}$ 
where addition/subtraction are understood modulo $N$. In the absence of flux, the occupation is conjugated under the dynamics, $n_\mathrm{PF}\rightarrow -n_\mathrm{PF}$.

This admits a natural interpretation in terms of $\mathbb{Z}_N$ anyons. We associate the eigenvalues of the string $S_{\curvearrowright}$ and the plaquette operator $W_p$ with a parafermionic anyon 
$\psi$ and a magnetic flux $m$, respectively. An anyon is then labeled by
$\psi^{n_\mathrm{PF}}\times m^p$. Introducing the electric anyon 
$e=\psi\times m$, the anyons may equivalently be written as $e^a\times m^b$ with $a,b\in\mathbb{Z}_N$.

The Floquet evolution then corresponds to the anyon transmutation $\psi^{n_\mathrm{PF}}\times m^p\leftrightarrow \psi^{p-n_\mathrm{PF}}\times m^p$ \cite{po_radical_2017}.
In terms of electric and magnetic charges, this takes the symmetric form
\begin{align}
    e^a \times m^b \leftrightarrow e^b \times m^a,
\label{eq:anyon transmut}
\end{align}
exchanging the electric and magnetic quantum numbers. This generalizes the result
from the $\mathbb{Z}_2$-case to arbitrary $\mathbb{Z}_N$. There are $N(N-1)/2$ period-two transmutation pairs ($a\neq b$), while $N$ anyons remain invariant under driving ($a=b$). This period-two anyon transmutation realizes a $\mathbb{Z}_N$ anyonic time crystal, generalizing discrete time crystals in $\mathbb{Z}_N$ clock models \cite{Surace2019,goss_qutrit_2026}.

Similar to Ref.~\cite{will_probing_2025}, we propose the following protocol to probe anyon transmutation on a qudit device, see Fig.~\ref{fig:fig2}b. (I) Initialize the system in the flux-free state defined by
\begin{align}
W_p\ket{\Psi_\mathrm{FF}}=\ket{\Psi_\mathrm{FF}},
\quad T^z_i T^z_j \ket{\Psi_\mathrm{FF}}=\omega^{N_{ij}}\ket{\Psi_\mathrm{FF}}
\label{eq:FF state}
\end{align}
where $N_{ij}\in\mathbb{Z}_N$ is the occupation of the parafermion pair on the $z$-bond $\langle i,j\rangle$. 
This corresponds to the
groundstate of the $\mathbb{Z}_N$ toric code, which can be prepared by a unitary circuit of depth linear in the system width \cite{iqbal_qutrit_2025}. 
(II) Locally apply bond operators to rearrange the parafermion strings with occupation $N_{ij}$ into half-plaquette pairings, leaving the gauge flux unchanged and resulting in the state $\ket{\Psi_0}$. (III) Create an anyon-antiparticle pair, and move one anyon to the boundary of the system by applying a string operator. The resulting state $\ket{\Psi_{e^{a}\times m^b}}$ serves as the initial state for the subsequent Floquet evolution. (IV) Probe the type of the remaining anyon by measuring a loop operator $\mathcal{L}$ encircling it. $\mathcal{L}$ creates a magnetic-anyon pair, moves one around the to-be-enclosed anyon, and subsequently annihilates the pair, see Fig.~\ref{fig:fig2}c.

To characterize the anyon transmutation, we define a non-equilibrium loop order parameter \cite{will_probing_2025}, with complex values,
\begin{align}\eta(t)=\frac{\langle\Psi_{e^a\times m^b}|{\mathcal{L}(t)|\Psi_{e^a\times m^b}\rangle }} 
    { \langle\Psi_0|\mathcal{L}(t)|\Psi_0\rangle},
\end{align}
where $\mathcal{L}(t)=(U_F^\dagger)^t \mathcal{L} ({U_F})^t$ and number of Floquet cycles $t$. In the limit of large loop diameters and an infinite system-size, it serves as a topological invariant. At the Clifford point the denominator is one, and thus the order parameter reduces to the expectation value of the time evolved loop operator.

As an example, we simulate the protocol for the $\mathbb{Z}_3$ case at $J=1$ using stabilizer simulations, see Fig.~\ref{fig:fig2}d and App.~\ref{SI:stabilizer_simulations} for details. The initial flux-free state is a stabilizer state, and both the preparation and Floquet protocols consist of Clifford gates.
 
The $\mathbb{Z}_3$ theory hosts nine anyon types, namely the vacuum, the electric and magnetic anyons $e$ and $m$, the parafermion $\psi$, the dyon $d = e\times m$ as well as their antiparticles, see Fig.~\ref{fig:fig2}a. The order parameter directly tracks the transmutation of these anyons under the Floquet dynamics. Its initial phase is determined by the electric quantum number $\eta(0)=\omega^a$. Its magnitude remains fixed at $|\eta(t)|=1$ throughout the evolution. For six of the nine anyon types, the phase alternates between two values, revealing three pairs of mutually transmuting anyons. Those pairs are mirrored with respect to the diagonal of Fig.~\ref{fig:fig2}d. In contrast, the vacuum, the dyon and its antiparticle remain invariant and are shown on the diagonal of Fig.~\ref{fig:fig2}d. Our simulation results are  consistent with the transmutation rule in Eq.~\eqref{eq:anyon transmut}. Away from $J=1$, we anticipate persistent oscillations of $\eta$, as in the $\mathbb{Z}_2$ case \cite{will_probing_2025}, signaling a prethermal regime in which anyon transmutation, and hence FTO, persists.

\section{Parafermionic edge modes}
\subsection{Correlation functions}
\begin{figure}[t]
    \centering
    \includegraphics[width=\linewidth]{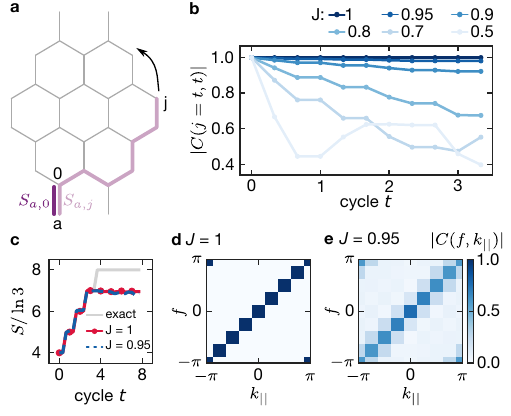}
    \caption{\textbf{Chiral parafermionic edge dynamics.} \textbf{a} Setup for measuring the edge correlation function using an auxiliary qudit as a fixed endpoint of the string. \textbf{b} Real space edge correlation function obtained from MPS simulations. At $J=1$ the signal remains one, 
    whereas for $J<1$ it
    decays over time. \textbf{c} Maximal entanglement entropy over all bipartitions along the MPS ordering (Fig.~\ref{fig:SI_fig_mps_snake}c) for $J=1$ (red) and $J=0.95$ (blue), starting from the stabilizer state shown in Fig.~\ref{fig:SI_fig_mps_snake}d. Gray: exact $J=1$ result from the stabilizer simulation. \textbf{d} Momentum-frequency space correlation function at $J=1$ shows the linearly dispersing chiral edge mode and its winding across the Floquet quasienergy Brillouin zone. \textbf{e} 
    Same for $J=0.95$. Despite broadening,
    the quasienergy winding of the chiral edge mode persists. All  data is obtained from MPS simulations with bond dimension
     $\chi = 3^7$.}
    \label{fig:fig3}
\end{figure}

To probe the chiral parafermion transport along the edge, we introduce an auxiliary qudit outside the system that does not participate in the Floquet evolution, see Fig.~\ref{fig:fig3}a. 
We define the correlator
\begin{align}
C(j,t)=\langle \Psi_0|S_{a,j}^\dagger(t)S_{a,0}(0)|{\Psi_0}\rangle,
\end{align}
where $j$ labels the sites along the boundary and $a$ the auxiliary site. Since $a$ is not evolved, its parafermionic contributions cancel between the two clock strings. Up to a conserved gauge string, the remaining operator therefore reduces to a two-point parafermion correlator,
\begin{align}C(j,t)\propto\langle\Psi_0|\beta^\dagger_j(t)\alpha_0(0)|{\Psi_0}\rangle.
\end{align}
Thus $C(j,t)$ directly tracks the propagation of a parafermion $\beta^\dagger_j(t)$ initially created at edge site $0$ to site $j$ after $t$ Floquet cycles.

At the Clifford point $J=1$, the drive translates a parafermion by one unit cell along the edge per Floquet period, resulting in ballistic chiral propagation, with $|C(j=t)| = 1$, see Fig.~\ref{fig:fig3}b. In momentum-frequency space, this yields $C(k_\parallel,f)\propto\delta(f-k_\parallel)$, corresponding to a linearly dispersing chiral edge mode, see Fig.~\ref{fig:fig3}d. The associated parafermion dispersion winds nontrivially across the quasienergy Brillouin zone.

To investigate the stability of the edge dynamics away from the Clifford point, we employ MPS simulations, using the TeNPy library~\cite{tenpy}, for details see App.~\ref{app:mps}. Rapid entanglement growth limits the accessible simulation times, see Fig.~\ref{fig:fig3}c. Already at the Clifford point an exact MPS representation of the stabilizer dynamics requires a bond dimension of up to $\chi = 3^8$, for the shown system size with our chosen MPS ordering, see App.~\ref{app:mps}, making simulations particularly challenging.
Nevertheless, we can reliably access the early-time edge dynamics. Away from the Clifford point, the model becomes generically non-integrable, causing two-point correlations to spread. Indeed, for $J<1$ we observe a decay and broadening of the edge correlations, as shown in Fig.~\ref{fig:fig3}b. Despite this, the characteristic quasienergy winding persists at $J=0.95$ in the Fourier-transformed correlation function, see Fig.~\ref{fig:fig3}e. Whether the response remains stable at long times and in the thermodynamic limit remains an important open question.

\subsection{Quantum information transport}
\begin{figure}[t]
    \centering
\includegraphics[width=\linewidth]{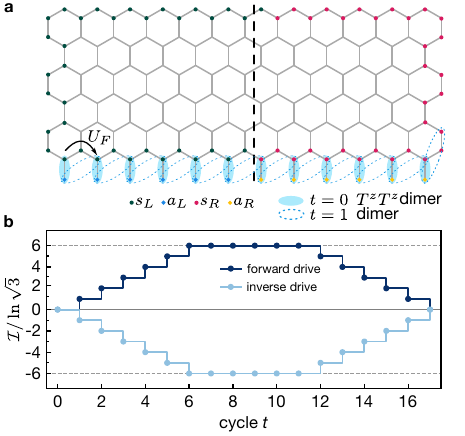}
    \caption{\textbf{Chiral mutual information.} 
    \textbf{a} Static auxiliary qutrits (square markers) 
    are attached to every second site along the bottom boundary, each initially prepared with the adjacent edge system qutrit in a $T^zT^z$ pair (lightblue shading). After one Floquet drive the dimer evolves into the blue dashed configuration. The black dashed line partitions the boundary into a left region $s_L$ (blue circles) and a right region $s_R$ (red circles). 
    \textbf{b} Chiral mutual information $\mathcal{I}$ vs. Floquet cycles. The growth of $\ln\sqrt{3}$ per period persists for the first six cycles, until the partners of all left auxiliary qutrits have crossed the cut, followed by a plateau and a linear decrease as the partners of the right auxiliaries cross the upper cut. Light blue: reversed drive $U_F^\dagger$. Data obtained with stabilizer simulations.}
    \label{fig:fig4}
\end{figure}

The parafermionic edge modes transport a quantized amount of quantum information, which we characterize using the chiral mutual information of Ref.~\cite{Duschatko2018}. Such quantized edge information flow is characterized by the chiral unitary index and related to a bulk invariant of two-dimensional Floquet topology
~\cite{po_radical_2017,zhang_bulkboundary_2023}. We probe the information edge flow by attaching auxiliary qutrits to the bottom of the system, and preparing each auxiliary $a_i$ in a $T^z_{s_i}T^z_{a_i}$ pair with the adjacent qutrit $s_i$ of the system, see Fig.~\ref{fig:fig4}a marked in blue. We then evolve only the physical qutrits of the system under $U_F$, while the auxiliaries remain static. 
After one Floquet period, the driven partner of each initial dimer is translated to the right along the edge, evolving the blue dimers in Fig.~\ref{fig:fig4}a into the dashed, blue configuration. Consequently, the system half of one entangled dimer is transported across the cut per Floquet period.

To measure the transfer of quantum information, we partition the boundary strip $s$ and the auxiliaries $a$ into left and right regions, $s_{L/R}$ and $a_{L/R}$ and define the chiral mutual information as
\begin{align}
\mathcal{I}(t) &= I(a_L,s_R)-I(a_R,s_L),\\
I(A,B)&=\frac{1}{2}\left[S(A)+S(B)-S(A\cup B)\right],
\end{align}
with the von Neumann entropy $S(A)=-\text{Tr}[\rho_A\ln\rho_A]$. 

Our stabilizer simulations at $J=1$ show that $\mathcal{I}$ increases by $\ln\sqrt{3}$ per Floquet period, see Fig.~\ref{fig:fig4}b. The information transported per period thus equals the radical chiral unitary index $\nu=\ln\sqrt{3}$~\cite{po_radical_2017, gross_index_2012}, which directly yields the fractionalized quantum dimension $\sqrt{3}$, \footnote{We note that Ref.~\cite{Duschatko2018} prescribes evaluating the transport over two Floquet periods for radical phases; at the Clifford point, where bulk and edge evolution decouple exactly, the per-period value is unambiguous.}.
The linear growth persists until the partners of all left auxiliaries have crossed the cut, after which $\mathcal{I}$ saturates. Once the partners of the right auxiliaries $a_R$ cross the cut at the top of the system, $\mathcal{I}$ decreases back towards zero. Reversing the drive, $U_F \rightarrow U_F^\dagger$, reverses the sign of $\mathcal{I}$, directly confirming the chirality of the quantized information transport, see Fig.~\ref{fig:fig4}b. While we explicitly show results for the $\mathbb{Z}_{N=3}$ case here, the setup can directly be generalized to arbitrary $N$.

\subsection{Parafermionic exchange statistics} 
The Floquet setup provides a natural platform to probe the exchange statistics of parafermions. Consider two parafermions $\alpha_{1,2}$ at the edge with charge $q_{12}=(\alpha_1\alpha_2)^\dagger=\omega^{n_\mathrm{PF}}$. For $\mathbb Z_3$, their exchange is generated by $\mathcal{B}_{12}=e^{-i\frac{2\pi}{9}(q_{12}+q_{12}^\dagger)}$ \cite{lindner_fractionalizing_2012}. This yields a relative phase $\phi_{0}=-4\pi/9$ for $n_\mathrm{PF}=0$ and $\phi_{1,2}=2\pi/9$ for $n_\mathrm{PF}=1,2$.

\begin{figure}[t]
    \centering
    \includegraphics[width=\linewidth]{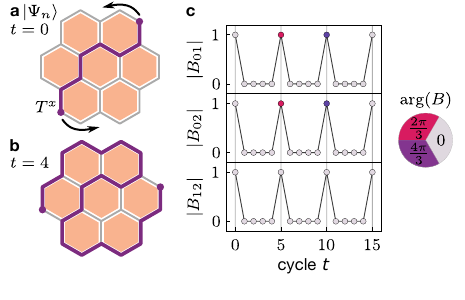}
    \caption{\textbf{Edge mode interferometry.} \textbf{a} The initial state $\ket{\Psi_n}$ contains a stretched string across the system, whose ends carry a pair of parafermions.
    Application of a local clock $T^x$ operator at one of its ends controls the occupation of the pair. For clarity, other strings and parafermion pairs are not shown. \textbf{b} The Floquet dynamics exchanges the two parafermions at the ends of the string. \textbf{c} The relative braiding phase is measured by comparing states with different occupations $n$ and $n^\prime$ at $J=1$. The string returns to its initial position after $t=5$ cycles, with relative phase of $\Delta \phi =2\pi/3$ or zero. Data obtained with stabilizer simulations.}
    \label{fig:fig5}
\end{figure}

We extract the exchange phase using the parafermion edge modes at $J=1$. Starting from the flux-free state $|\Psi_\mathrm{FF}\rangle$, we prepare a nonlocal parafermion pair at opposite edges of the system by stretching a clock string across the system, resulting in $|\Psi_{n}\rangle$ with $n\equiv n_\mathrm{PF}$, see Fig.~\ref{fig:fig5}a. The occupation is controlled by a local clock operation $T^x$ at one string endpoint. The chiral Floquet dynamics along the boundary then exchanges the parafermions after a system-size-dependent number of Floquet periods, see Fig.~\ref{fig:fig5}b. The relative braiding phase between sectors $n,n^\prime$ can be extracted by
\begin{align}
B_{nn^\prime}=\langle \Psi_{n}|
\left(T^x_{0}\right)^{n-n^\prime}(t)
 \left(T^x_{0}\right)^{n^\prime-n}(0)|
\Psi_{n}\rangle.
\end{align}
At the Clifford point, our stabilizer simulations show that $B_{nn'}(t)$ is nonzero only when the stretched string has encircled the boundary and returned to its initial configuration, see Fig.~\ref{fig:fig5}c. At these revival times, its phase yields the relative braiding phase. We find $\phi_{1,2}-\phi_0=2\pi/3$ for $n=0$ and $n'=1,2$, while $\phi_2-\phi_1=0$ for $n=1$ and $n'=2$, consistent with our expectation. The Floquet protocol thus converts parafermionic exchange statistics into an interferometric signal accessible through local qudit operations and time evolution, without adiabatic braiding.

\section{Summary and Outlook}
We have shown that the generalized Kitaev model built from clock matrices provides a versatile testbed for studying $\mathbb{Z}_N$ FTO. A Fradkin-Kadanoff type mapping reveals the underlying parafermionic nature of the model and offers a natural framework for understanding its characteristic dynamical phenomena. These include chiral parafermionic edge transport, fractional information transport, parafermion braiding and anyon transmutation in the bulk. Our drive admits a Clifford point for all $\mathbb{Z}_N$, at which these signatures can be simulated efficiently using stabilizer simulations, as we have demonstrated explicitly for $\mathbb{Z}_3$. Away from it, MPS simulations allowed us to probe the dynamics over the first few driving cycles, before the rapid growth of entanglement hindered convergence.
Accessing $\mathbb{Z}_N$ FTO beyond these classically tractable time scales will require complementary experimental and numerical approaches.
Experimental realizations, for instance on quantum simulators with native qudit operations, offer a promising route to explore the Floquet dynamics at larger system sizes and longer times.
On the numerical side, methods tailored to near-Clifford dynamics, such as sparse Pauli propagation~\cite{rall_pauli_2019,rudolph2026paulipropagation,begusic2024utility,begusic2025spd,begusic2025cpt} or Clifford-augmented tensor-network approaches~\cite{qian2024camps,qian2025catdvp,mello2025clifforddressed}, may extend the accessible regime, and first generalizations from qubits to qudits have been developed~\cite{harper2026gcamps}.

More broadly, an interesting direction is to extend the present construction to realize higher-period anyon permutations. While our drive realizes period-two exchange for any $N$, the richer anyon structure of $\mathbb{Z}_N$ for larger $N$ also permits longer permutation-cycles with no $\mathbb{Z}_2$ counterpart. Developing experimentally accessible Floquet protocols for such dynamics and characterizing their associated edge transport would broaden the phenomena accessible in driven qudit systems.

\textit{Code availability. ---} Code is available on Zenodo upon reasonable request \cite{zenodo}.

\textit{Use of AI tools. ---} Claude Code (Opus 5.5, Fable 5.1) was used to assist with algebraic consistency checks, code development, literature review, and language editing. All AI-assisted material was reviewed and revised by the authors, who take full responsibility for the content.

\begin{acknowledgments}
We thank B.~Jobst, R.~M.~Yepes, K.~Shtengel, C.F.B.~Lo and A.~Vishwanath for fruitful discussions. M.W. was supported by a scholarship of the German Academic Exchange Service (DAAD). A.G-S. was funded by UK Research and Innovation (UKRI) under the UK government’s Horizon Europe funding guarantee [grant number EP/Y036069/1] and acknowledges support from the Leverhulme Trust through Research Project Grant [RPG-2024-112]. We acknowledge support from the Deutsche Forschungsgemeinschaft (DFG, German Research Foundation) under Germany’s Excellence Strategy--EXC--2111--390814868, TRR 360 – 492547816 and DFG grants No. KN1254/1-2, KN1254/2-1, the European Union (grant agreement No 101169765), as well as the Munich Quantum Valley, which is supported by the Bavarian state government with funds from the Hightech Agenda Bayern Plus.  M.W. and F.P. acknowledge
support from the DFG Research Unit FOR 5522 (project id 499180199). 
\end{acknowledgments}

\appendix
\setcounter{secnumdepth}{2}
\counterwithout{equation}{section}  
\setcounter{figure}{0}
\setcounter{equation}{0}
\setcounter{table}{0}
\renewcommand{\thefigure}{S\arabic{figure}}
\renewcommand{\theequation}{S\arabic{equation}}
\renewcommand{\thetable}{S\arabic{table}}

\section{Parafermion mapping}
\label{SI:parafermion_mapping}
Here we provide additional details on the parafermion mapping. 
Let us first recall their definitions:
\begin{align}
A_{r}&=\Fop_{r}T^x_{r},
\qquad
&\widetilde A_{r}&=\Fop_{r}{T^x_{r}}^\dagger, \\
B_{r}&=\Fop_{r}{T^z_{r}}^\dagger {T^x_{r}}^\dagger, 
\qquad
&\widetilde B_{r}&=\Fop_{r}T^x_{r} T^z_{r},
\label{eq:parafermions}
\end{align}
where we have defined 
\begin{align}
    \Fop_r&=\prod_{r' \prec r}\left(T^z_{r'}\right)^{s_{r'}},    
    \quad s_{r'} =
    \begin{cases}
        +1, & r' \in \text{white},\\
        -1, & r' \in \text{black}.
    \end{cases}
    \label{eq:app_string}
\end{align}
Here, $r'\prec r$ denotes that $r'$ precedes $r$ along the chosen row-by-row ordering, see Fig.~\ref{fig:fig1}c. Thus, $\Fop_r$ contains $T^z$ on every preceding white site and ${T^{z}}^\dagger$ on every preceding black site. 
\subsection{Same-site algebra}
At same sites, their commutation relations are determined by the clock operator at site $r$, since all of them share the same string $\Fop_r$. Therefore, we find:

\begin{align}
A_{r}\widetilde A_{r} &= \widetilde A_{r} A_{r} \, \leftrightarrow \, T^x(T^x)^\dagger = (T^x)^\dagger T^x\\
B_{r}\widetilde B_{r} &= \widetilde B_{r} B_{r} \leftrightarrow  T^y(T^y)^\dagger= (T^y)^\dagger T^y\\
A_{r}\widetilde B_{r} &= \om^{-1} \widetilde B_{r} A_{r} \leftrightarrow T^x(T^y)^\dagger = \om^{-1} (T^y)^\dagger T^x\\
\widetilde A_{r}B_{r} &= \om^{-1}  B_{r} \widetilde A_{r} \leftrightarrow (T^x)^\dagger T^y = \om^{-1} T^y (T^x)^\dagger \\
A_{r}B_{r} &=\om\, B_{r}A_{r} \leftrightarrow T^xT^y = \om T^y T^x\\
\widetilde A_{r}\widetilde B_{r}
&=\om\,\widetilde B_{r}\widetilde A_{r} \leftrightarrow (T^x)^\dagger (T^y)^\dagger = \om (T^y)^\dagger (T^x)^\dagger
\label{eq:commute_samesite}
\end{align}

\subsection{Off-site algebra}
Let us now consider exchange statistics on different sites $r$ and $r'$, where $r< r'$ in the row-by-row ordering, shown in Fig.~\ref{fig:fig1}c. 
We then find that if $r$ is a white site 
\begin{align}
    A_r\alpha_{r'} &= \om^{-1} \alpha_{r'} A_r, \,
    &\widetilde B_r\alpha_{r'} &= \om^{-1} \alpha_{r'} \widetilde B_r, \,  \\
    \widetilde A_r\alpha_{r'} &= \om \alpha_{r'} \widetilde A_r,  \,
     &B_r\alpha_{r'} &= \om \alpha_{r'}  B_r,
\end{align}
and if $r$ is a black site
\begin{align}
    A_r\alpha_{r'} &= \om \alpha_{r'} A_r, \,
    &\widetilde B_r\alpha_{r'} &= \om \alpha_{r'} \widetilde B_r, \,  \\
    \widetilde A_r\alpha_{r'} &= \om^{-1} \alpha_{r'} \widetilde A_r,  \,
     &B_r\alpha_{r'} &= \om^{-1} \alpha_{r'}  B_r,
\end{align}
with $\alpha_{r'} \in (A_{r'},\widetilde{A}_{r'},B_{r'},\widetilde{B}_{r'})$.
Those offsite commutation relations are consistent with parafermionic off-site exchange relations, when interpreted in the order
\begin{align*}
&\widetilde B_1 \rightarrow A_1 \rightarrow B_2\rightarrow\widetilde A_2\rightarrow\ldots \rightarrow B_L\rightarrow\widetilde A_L \\
&\rightarrow\widetilde B_L\rightarrow A_L\rightarrow \ldots \rightarrow\widetilde B_2\rightarrow A_2 \rightarrow B_1\rightarrow \widetilde A_1,
\end{align*}
where the first physical site is chosen to lie on the black sublattice and the last one on the white sublattice. Here, $1,\ldots,L$ label physical sites along the spatial row-by-row ordering. On distinct physical sites, the parafermions satisfy the standard $\mathbb Z_N$ exchange relation
\begin{align}
\alpha \beta=\omega\beta\alpha,
\qquad
\alpha,\beta\in\{A,B,\widetilde A,\widetilde B\},
\end{align}
for $\alpha$ preceding $\beta$ in the parafermionic snake ordering.
\subsection{Bond identities}
Let us now rewrite the bonds of the drive in terms of these constrained parafermions.
The $x$-bond becomes
\begin{align}
\widetilde B_{r} \widetilde A_{r+1}^\dagger = \Fop_r T_r^x T_r^z T_{r+1}^x \Fop_{r+1}^\dagger = T^x_{r} T^x_{r+1} 
\end{align}
where we have used $\mathcal F_{r+1}=\mathcal F_rT^z_r$, since the lower site of every $x$-bond belongs to the white sublattice.
Analogously we find for the $y$-bond:
\begin{align}
    A_{r}^\dagger B_{r+1} &= (T^x_r)^\dagger \Fop_{r}^\dagger \Fop_{r+1} (T_{r+1}^z)^\dagger(T_{r+1}^x)^\dagger  \\ 
    &= (T^x_r)^\dagger (T_{r+1}^z)^\dagger(T_{r}^z)^\dagger(T_{r+1}^x)^\dagger \\&= \om \zeta^{-2}T^y_{r}T^y_{r+1}.
\end{align}
The second line uses $\mathcal F_r^\dagger\mathcal F_{r+1}=(T^z_r)^\dagger$, which holds because the lower site of every $y$-bond is black.
For the vertical $z$-bonds, we first note the on-site identities
\begin{align}
T^z_r = \widetilde A_r B_r^\dagger = \om\, B_r^\dagger \widetilde A_r ,
\end{align}
in which the strings cancel because one factor carries $\Fop_r$ and the other $\Fop_r^\dagger$.
Using $T^z_r=\widetilde A_r B_r^\dagger$ at the lower site, $T^z_{r+L_x}=\om B_{r+L_x}^\dagger\widetilde A_{r+L_x}$ at the upper site, and the fact that $\widetilde A_{r+L_x}$ commutes with $T^z_r$, the $z$-bond becomes
\begin{align}
T^z_{r+L_x} T^z_r
&= \om\, B_{r+L_x}^\dagger \widetilde A_{r+L_x}\, \widetilde A_r B_r^\dagger\\
&= B_{r+L_x}^\dagger \widetilde A_ru_r,
\end{align}
where we have defined
\begin{align}
u_r = \om\, B_r^\dagger \widetilde A_{r+L_x}.
\end{align}

\begin{figure}[t]
    \centering
    \includegraphics[width=\linewidth]{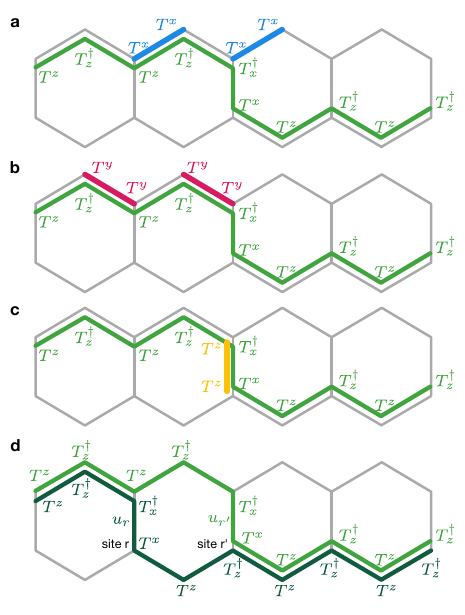}
\caption{\textbf{Intersections of gauge links.} \textbf{a--c} Possible intersections of $x,y,z$-bonds with the gauge string $u_r$. All terms in the Floquet operator commute with $u_r$. \textbf{d} Two intersecting gauge links do not commute locally at sites $r+L_x$ and $r'$. Since the two shared sites will pick up opposite phases, different gauge links commute overall.}
\label{fig:SI_fig_gauge_link_commute}
\end{figure}

Next, we will argue that terms $u_{r}$ actually commute with the Floquet operator and with each other, and are therefore conserved $\mathbb Z_N$ gauge links. 
A gauge link $u_{r}$ is the product of $T_r^x(T_{r+L_x}^x)^\dagger$ as well as $T^z$ on white sites and $(T^z)^\dagger$ operators on all black sites that are in between site $r$ and $r+L_x$ in the row-by-row ordering, see Fig.~\ref{fig:fig1}d.
We inspect all bonds appearing in the pulses $U^\alpha$ case-by-case.
Fig.~\ref{fig:SI_fig_gauge_link_commute} shows different intersections between bonds and the gauge string $u_r$.
For the $x$-bond we have the following two possibilities:
\begin{align}
    T^x_{r}T^x_{r+1}T^z_r (T^z_{r+1})^\dagger &= T^z_r (T^z_{r+1})^\dagger T^x_{r}T^x_{r+1}\\
    T^x_{r}T^x_{r+1} \mathbbm{1}_{r}T^x_{r+1} &= \mathbbm{1}_{r}T^x_{r+1}  T^x_{r}T^x_{r+1},
\end{align}
see Fig.~\ref{fig:SI_fig_gauge_link_commute}a.
For the $y$-bond we need to verify:
\begin{align}
    T^y_{r}T^y_{r+1}(T^z_{r})^\dagger T^z_{r+1} &= (T^z_{r})^\dagger T^z_{r+1}T^y_{r}T^y_{r+1}\\
    T^y_{r}T^y_{r+1}(T^z_{r})^\dagger (T^x_{r+1})^\dagger &= (T^z_{r})^\dagger (T^x_{r+1})^\dagger T^y_{r}T^y_{r+1}
\end{align}
as shown in Fig.~\ref{fig:SI_fig_gauge_link_commute}b.
For the $z$-bond we have:
\begin{align}
    T^z_{r}T^z_{r+L_x}T^x_r (T^x_{r+L_x})^\dagger &= T^x_r (T^x_{r+L_x})^\dagger T^z_{r}T^z_{r+L_x},
\end{align}
see Fig.~\ref{fig:SI_fig_gauge_link_commute}c.
To show that $u_{r}$ are indeed conserved $\mathbb Z_N$ gauge links we also need to show that they commute with each other. Considering two gauge links $u_{r}$ and $u_{r'}$ they can either not share any sites and do therefore commute trivially or intersect with each other. In the latter case we will now argue that they also commute. For this we would like to discuss the example where they are situated in the same row as shown in Fig.~\ref{fig:SI_fig_gauge_link_commute}d. One finds, that they do commute on all sites except at site $r'$ and $r+L_x$. On site $r+L_x$ one encounters the commutation relation
\begin{align}
    (T^x)^\dagger T^z = \om\, T^z (T^x)^\dagger ,
\end{align}
whereas on site $r'$
\begin{align}
    (T^z)^\dagger T^x = \om^{-1} T^x(T^z)^\dagger.
\end{align}
Both phases are for moving $u_{r'}$ through $u_r$.
We therefore find, that the phases accumulated on those two sites perfectly cancel out and therefore two gauge strings commute with each other. It can be checked that any other intersection of two strings will yield the same situation. 
Therefore, we have shown
\begin{align}
    [U_F,u_{r}] &= 0,\\
    [u_{r},u_{r'}] &= 0.
\end{align}
Products of neighboring gauge links reproduce the plaquette flux,
\begin{align}
W_p=\zeta^4u_r^\dagger u_{r+2},
\label{eq:SI_plaquette_gauge}
\end{align}
as illustrated in Fig.~\ref{fig:SI_parafermion_mapping}a. The gauge variables therefore encode the static $\mathbb{Z}_N$ flux sector of the model.

\subsection{Mapping strings to parafermions}
\begin{figure}[t]
    \centering
    \includegraphics[width=\linewidth]{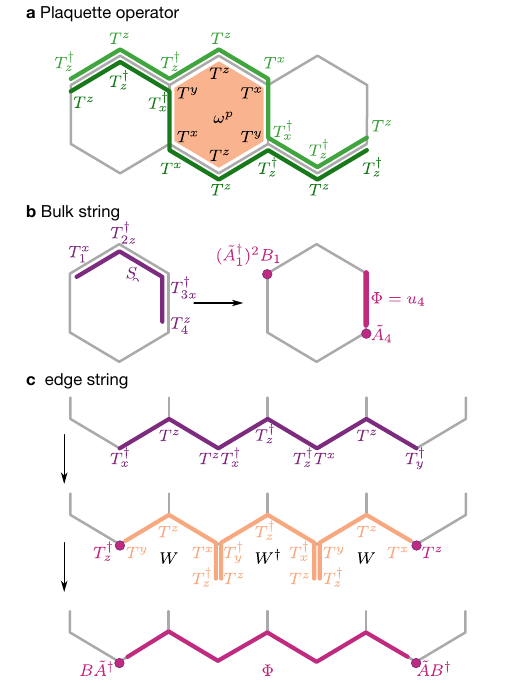}
    \caption{\textbf{Mapping the strings to parafermions} \textbf{a} Decomposition of the plaquette operator $W_p$ into gauge strings. \textbf{b} In the bulk, a clock string across one plaquette maps onto a parafermion bilinear at its endpoints, accompanied by a gauge string. \textbf{c} At the boundary, clock strings likewise map onto parafermion bilinears and an invariant string. The latter can be expressed as a product of plaquette operators along the edge.} \label{fig:SI_parafermion_mapping}
\end{figure}

Next, we provide additional details on the relation between clock strings and the parafermionic representation. We first consider a connected clock string 
\begin{align}
S_{rr'} = \prod_{\langle l,k\rangle\in\gamma_{r,r'}}K_{lk}
\end{align}
along a path $\gamma_{rr'}$. The string takes the form
\begin{align}
    S_{rr'} = \alpha_r\beta_{r'}\Phi_{\gamma_{rr'}},
\label{eq:SI_string_mapping}
\end{align}
where $\alpha_r$ and $\beta_{r'}$ are parafermions at the two endpoints and $\Phi_{\gamma_{r,r'}}$ is a product of conserved quantities along the path. Note that we have absorbed all the phases into the gauge string $\Phi_{\gamma_{rr'}}$. For example, the clock strings across a plaquette with qudit labels $1,\dots,6$ in the bulk,
\begin{align}
S_{\curvearrowright}=\zeta^2T^x_1  (T^z_2)^\dagger (T^x_3)^\dagger T^z_4
\end{align} 
can be expressed in parafermionic operators, 
\begin{align}
S_{\curvearrowright}=\zeta^2\om^{-2}(\tilde{A}_1^\dagger)^2B_1 \tilde{A}_4 u_{4},
\end{align}
see Fig.~\ref{fig:SI_parafermion_mapping}b. We note that $(\tilde{A}^\dagger_1)^2B_1$ also behaves as a parafermion with respect to all the other parafermions.

The same construction applies to a string on the edge, see Fig.~\ref{fig:SI_parafermion_mapping}c. Again the string maps onto endpoint parafermions multiplied by a conserved string. The string connecting the two parafermions can be expressed as a product of plaquette operators of the edge, see Fig.~\ref{fig:SI_parafermion_mapping}c. Consequently, it commutes with the drive and the nontrivial dynamics solely arise from the motion of the parafermionic endpoints.

\subsection{Parafermion dynamics}
\label{sec:pf dynamics}
At the Clifford point, the parafermion dynamics follows directly from the braid operator \eqref{eq:braid operator}. We find the braid rules under $U^{\alpha\dagger}(\cdot)U^\alpha$ for the different bonds in Eqs.~\eqref{eq:xx-bond PF}-\eqref{eq:zz-bond PF},
\begin{align}
x:&\quad \widetilde B_r\to e^{i\pi\frac{N-1}{N}}\,\widetilde A_{r+1},\\
&\quad \widetilde A_{r+1}\to e^{i\pi\frac{N-1}{N}}\,\widetilde A_{r+1}\widetilde B_r^\dagger\widetilde A_{r+1},\\
y:&\quad B_{r+1}\to e^{i\pi\frac{N+3}{N}}\zeta^{-2}A_r,\\
&\quad A_r\to e^{i\pi\frac{N+3}{N}}\zeta^{-2}A_rB_{r+1}^\dagger A_r,\\
z:&\quad \widetilde A_r\to e^{i\pi\frac{N+1}{N}}\,B_{r+L_x}u_r^\dagger,\\
&\quad B_{r+L_x}\to e^{i\pi\frac{N+1}{N}}\,B_{r+L_x}u_r^\dagger\widetilde A_r^\dagger B_{r+L_x}.
\end{align}
The first rule of each pair is a free hop, from the white to the black site on $x$- and $y$-bonds and from the lower
to the upper site on $z$-bonds, where it carries the gauge link. The second rule is the hop back dressed by the bond
charge $K_{kl}=\omega^q$, e.g.\ $\widetilde A_{r+1}\widetilde B_r^\dagger\widetilde A_{r+1}\propto\widetilde B_r(K^x)^{-2}$, which acquires the phase $\omega^{-2q}$. The reversed drive exchanges the two directions.

\section{Exactly solvable point}

\subsection{Clifford rules}
\label{app:clifford}

We show that the Floquet pulses become Clifford gates at the sweet spot. We
first derive a polynomial expansion from the exponential representation of
the bond pulses,
\begin{equation}
U^{\alpha} = \prod_{\langle i,j\rangle_\alpha}e^{iJ\sum_{n=1}^{N-1} c_n[T_i^\alpha T_j^\alpha]^n}
= \prod_{\langle i,j\rangle_\alpha} \sum_{n=0}^{N-1} a_n\, (T_i^\alpha T_j^\alpha)^n.
\label{eq:U bonds}
\end{equation}
This expansion follows because the bond operators $K = T_i^\alpha T_j^\alpha$
satisfy $K^N=1$, so every function of $K$ can be expressed as a polynomial in
the basis $\{1,K,\dots,K^{N-1}\}$. 
Below, we explicitly derive the coefficients
\begin{equation}
a_n=\frac{e^{i\lambda}}{\sqrt N}\,e^{i\pi n(n+N)/N},
\label{eq:a_n}
\end{equation}
where 
\begin{align}
\lambda=\frac{\pi(N-1)(N-2)}{12N} 
\end{align}
is a global phase. 

We first illustrate the construction for the $\mathbb{Z}_3$ case. Here, the weights
are $c_1=c_2=-\frac{2\pi}9$. The bond pulse is
\begin{equation}
U_{ij}^{\alpha\alpha} = e^{-\frac{2\pi i}{9}J(K+K^\dagger)}
\end{equation}
The exponential can be computed using $(K+K^\dagger)^2=2+(K+K^\dagger)$.
Iterating, we find
\begin{equation}
(K+K^\dagger)^n = \alpha_n+\beta_n(K+K^\dagger)
\end{equation}
with $\alpha_n=\frac13(2^n+2(-1)^n)$ and $\beta_n=\frac13(2^n-(-1)^n)$.
Plugging this back into the exponential yields
\begin{equation}
U_{ij}^{\alpha\alpha} = \frac13 e^{\frac{2\pi i}{9}J}
\left[(\omega^{-J}+2)+(\omega^{-J}-1)(K+K^\dagger)\right],
\end{equation}
where $\omega=e^{2\pi i/3}$. 
At the sweet spot $J=1$,
this gives $a_0=\frac{e^{i\pi/18}}{\sqrt3}$ and $a_1=a_2=\frac{e^{i\pi/18}\,\omega^{2}}{\sqrt3}$.

The argument extends directly to arbitrary $\mathbb{Z}_N$. The coefficients are obtained from the discrete Fourier transform
\begin{equation}
a_n = \frac1N\sum_{k=0}^{N-1}\exp\Big[iJ\sum_{r=1}^{N-1}c_r\omega^{kr}\Big]\,\omega^{-kn}.
\label{eq:DFT}
\end{equation}
On the eigenspace $K=\omega^{k}$ the pulse acts as $e^{iJ\varphi_k}$ with
$\varphi_k=\sum_{r=1}^{N-1}c_r\omega^{kr}$. Using
$\sum_{r=1}^{N-1}\omega^{kr}\sin^{-2}(\pi r/N)=\frac{N^2-1}{3}-2k(N-k)$ for
$0\le k\le N$, one finds
\begin{equation}
\varphi_k=-\frac{\pi(N^2-1)}{6N}+\frac{\pi}{N}\,k(N-k),
\label{eq:phik}
\end{equation}
so that at $J=1$ the $k$-dependent part of $e^{i\varphi_k}$ is the quadratic
phase $e^{i\pi k(N-k)/N}$, which is the defining property of a Clifford gate
diagonal in $K$. Inserting Eq.~\eqref{eq:phik} into Eq.~\eqref{eq:DFT} and
completing the square turns the sum into a quadratic Gauss sum, with the result given in Eq.~\eqref{eq:a_n}.

The polynomial expansion allows us to derive the Clifford update rules. We
consider an arbitrary clock string on a bond $(i,j)$,
\begin{equation}
T_i^{x a_i}T_i^{z b_i}T_j^{x a_j}T_j^{z b_j} \equiv (a_i,b_i,a_j,b_j),
\label{eq:bond in ZN integers}
\end{equation}
with $\mathbb{Z}_N$-integers $a_i,b_i,a_j,b_j$. The action of the unitary takes the generic Clifford form
\begin{equation}
U_{ij}^{\alpha\alpha\dagger}\left(T_i^{xa_i}T_i^{zb_i}T_j^{xa_j}T_j^{zb_j}\right)U_{ij}^{\alpha\alpha}
= \omega^{\sigma}\,T_i^{xa_i'}T_i^{zb_i'}T_j^{xa_j'}T_j^{zb_j'},
\end{equation}
where the primed exponents and phase $\sigma$ depend on the bond type. We find the update rules
\begin{align}
\label{eq:clifford xx}
U_{ij}^{XX}:~
(a_i,b_i,a_j,b_j)
&\mapsto
(a_i-\mu,b_i,a_j-\mu,b_j),
\nonumber\\
\mu&=b_i+b_j,\nonumber\\
\sigma&=-\sigma_0,
\\[3pt]
\label{eq:clifford yy}
U_{ij}^{YY}:~
(a_i,b_i,a_j,b_j)
&\mapsto
(a_i+\mu,b_i+\mu,a_j+\mu,b_j+\mu),
\nonumber\\
\mu&=(a_i+a_j)-(b_i+b_j),
\nonumber\\
\sigma&=\sigma_0+(1-\epsilon_N)\,\mu+\mu(a_i+a_j),
\\[3pt]
\label{eq:clifford zz}
U_{ij}^{ZZ}:~
(a_i,b_i,a_j,b_j)
&\mapsto
(a_i,b_i-\mu,a_j,b_j-\mu),
\nonumber\\
\mu&=-(a_i+a_j),\nonumber
\\
\sigma&=\sigma_0,
\end{align}
where $\sigma_0=\mu(N+\mu)/2$, and $\epsilon_N\in\{0,1\}$ is defined by
$\zeta^{2}=\omega^{\epsilon_N}$, i.e.\ $\epsilon_N=0$ for odd and
$\epsilon_N=1$ for even $N$. It accounts for the phase $\zeta$ in the
definition of $T^y$, which enters the $y$-bond operator as
$K=T^y_iT^y_j=\zeta^{2}\,(T^z_iT^x_i)^\dagger(T^z_jT^x_j)^\dagger$ and hence
multiplies the phase of $U^\dagger M U$ by $\zeta^{-2\mu}$. For even $N$ the exponent $\sigma_0$ can be half-integer. In that case,
$\omega^{\sigma}$ is understood as $e^{2\pi i\sigma/N}$, so that the phases are
$2N$-th roots of unity, as is generally the case for the qudit Pauli group in
even dimensions~\cite{gottesman_fault_1999}.

We checked the rules numerically against the action of Eq.~\eqref{eq:U bonds} on every one of the $N^4$ possible input strings of Eq.~\eqref{eq:bond in ZN integers}, for all three bond types up to $N\leq 9$, by using the orthogonality relation
\begin{equation}
\mathrm{Tr}[(T^{xa}T^{zb})(T^{xa'}T^{zb'})^\dagger] = N\delta_{a,a'}\delta_{b,b'}.
\end{equation}

\begin{figure*}[t]
    \centering
    \includegraphics[width=\linewidth]{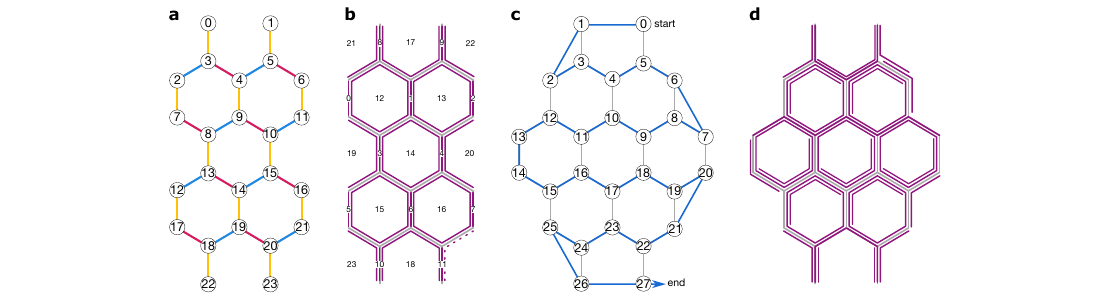}
    \caption{\textbf{Setup for stabilizer and MPS simulations.} \textbf{a} Site indices and bond types. \textbf{b} Flux-free state $|\Psi_\mathrm{FF}\rangle$: support of the stabilizer generators (purple). $0,\dots, 11$ are the $z$-bond operators and $12,\dots, 23$ (truncated) are the conserved plaquette operators. The dashed lower-right corner plaquette is not an independent generator: its flux is fixed by the others. \textbf{c} MPS snake ordering. \textbf{d} 
    Initial state $\ket{\Psi_0}$ of the Floquet evolution: support of the stabilizer generators (purple). It is used for simulating the edge correlation function.}
    \label{fig:SI_fig_mps_snake}
\end{figure*}

\subsection{Stabilizer simulations}
\label{SI:stabilizer_simulations}

The Clifford structure of the Floquet circuit enables an efficient stabilizer simulation. For literature on qudit stabilizers, see \cite{gottesman_fault_1999,spagnoli2026quditstabilisercodesmathbbzn,GHEORGHIU2014505,Farinholt_2014,2481591.2481597}. We perform these simulations for $N=3$, where $\mathbb{Z}_3$ is a finite field, so that standard Gaussian elimination can be used. This applies to any prime $N$, whereas composite $N$ requires more care since $\mathbb{Z}_N$ is then only a ring and Gaussian elimination needs to be generalized~\cite{Hostens2005,Brandl2026}.

For the $\mathbb{Z}_3$ Kitaev model with open boundary conditions and $N_q$ qudits, the flux-free state is a stabilizer state whose stabilizer group is generated by $\frac{N_q}{2}$ $T^z_iT^z_j$-bond operators on sites $i,j$ together with $\frac{N_q}{2}$ plaquette operators. We explicitly construct these generators and verify that they mutually commute by checking that all pairwise symplectic products vanish. Their linear independence is confirmed by Gaussian elimination 
over $\mathbb{Z}_3$, i.e. using arithmetic modulo three. 

The plaquette operators are defined by traversing each plaquette clockwise. Plaquette operators at the boundary are incomplete due to the open boundaries for the rectangular geometry used in the order parameter calculation, with armchair edges along the vertical direction and bowl edges along the horizontal direction. 
The support of the stabilizer generators for the flux-free state $\ket{\Psi_{\mathrm{FF}}}$ is shown in Fig.~\ref{fig:SI_fig_mps_snake}d.
For the ring geometry used in the correlation function calculation, one stabilizer wraps around the system and is reminiscent of the one-form symmetry present for periodic boundaries, see Fig.~\ref{fig:SI_fig_mps_snake}d for the dynamically prepared state $\ket{\Psi_{0}}$ (conserved stabilizers are the same as for $\ket{\Psi_\mathrm{FF}}$).

The preparation protocol described in the main text is implemented directly on the stabilizer generators. The local bond operations rearrange the initial $z$-pairings into the desired parafermion pairings while leaving all plaquette stabilizers invariant. We verified the implementation by comparing against exact state-vector simulations for small system sizes.

The Floquet dynamics are simulated using the Clifford update rules derived above. State evolution updates the stabilizer generators as $S_i\mapsto U_FS_iU_F^\dagger$. Equivalently, expectation values can be computed by evolving a generalized Pauli string as $O\mapsto U_F^\dagger OU_F$ while keeping the initial stabilizer state fixed. To evaluate the expectation value of an operator $O$, we first compute its symplectic products with the stabilizer generators. If $O$ does not commute with at least one stabilizer, then
\begin{align}
\langle\Psi|O|\Psi\rangle
&=
\langle\Psi|S_i^\dagger O S_i|\Psi\rangle
=
\omega^m\langle\Psi|O|\Psi\rangle,
\end{align}
with $m\neq0$, implying $\langle O\rangle=0$. Otherwise, $O$ commutes with the entire stabilizer group and therefore is proportional to an element of the stabilizer group.
Its decomposition into stabilizer generators, together with the associated phase, is obtained by Gaussian elimination over $\mathbb{Z}_3$, from which the expectation value follows immediately.

\section{Matrix-Product State Simulations}
\label{app:mps}

To study the Floquet dynamics away from the Clifford point we employ MPS simulations. These simulations are implemented using the Python library TeNPy~\cite{tenpy}.

To represent the two-dimensional system, shown in Fig.~\ref{fig:fig3}a, as an MPS we order the lattice sites into a suitable one-dimensional geometry. We start at the upper right site and move horizontally along $x$ and $y$ bonds until reaching the boundary of the system. Then we move to the next row and reverse the direction, see Fig.~\ref{fig:SI_fig_mps_snake}c. This way, sites sharing an $x$- or $y$-bond remain neighbors in the $1$D geometry, and only $z$-bonds become long-range. The initial state $|\Psi_0\rangle$, see Fig.~\ref{fig:SI_fig_mps_snake}d, is a stabilizer state. Rather than explicitly applying the full preparation circuit, we obtain $|\Psi_0\rangle$ using DMRG~\cite{PhysRevLett.69.2863,Schollwoeck2011, tenpy} by constructing a Hamiltonian whose mutually commuting terms are the stabilizer generators of the desired state. 
The Floquet evolution is implemented according to the three sequential bond pulses $U^x$, $U^y$ and $U^z$ and works 
in the representation obtained by the antiunitary map $\Theta=\mathcal{C}\mathcal{K}$, with $\mathcal{C}\ket{n}=\ket{-n}$ and $\mathcal{K}$ complex conjugation, under which $T^x\to T^{x\dagger}$, $T^z\to T^z$ and $U^\alpha\to\Theta U^\alpha\Theta^{-1}$. Since $\Theta$ is antiunitary, expectation values are complex conjugated, which leaves $|C(j,t)|$ and all entanglement entropies invariant. Since the $x$- and $y$-bonds connect neighboring sites in the chosen MPS ordering, the corresponding two-site gates can be applied using standard time-evolving block-decimation (TEBD) updates~\cite{Vidal2004}. We sweep from left to right for the $x$-pulse and from right to left for the $y$-pulse. In contrast, the $z$-pulses are then represented as matrix-product operators (MPO) acting between the corresponding sites. Applying this MPO increases the MPS bond dimension, after which the state is compressed using singular value decompositions while retaining the maximal bond dimensions. Away from the Clifford point, the Floquet evolution generates entanglement and the bond dimension required for an accurate MPS representation grows rapidly with time. This limits the simulations to relatively short evolution times and provides a direct measure of the increasing classical complexity away from the exactly tractable Clifford limit. 

\bibliographystyle{apsrev4-1}
\bibliography{bibliography}
\end{document}